\documentclass[11pt]{iopart}

\usepackage{iopams}
\usepackage{color}

\usepackage{graphicx}
\usepackage[colorlinks,bookmarks=false,citecolor=blue,linkcolor=red,urlcolor=blue]{hyperref}

\newcommand{\beq}{\begin{equation}}
\newcommand{\eeq}{\end{equation}}
\newcommand{\bma}{\begin{math}}
\newcommand{\ema}{\end{math}}
\newcommand{\beqa}{\begin{eqnarray}}
\newcommand{\eeqa}{\end{eqnarray}}

\def\opone{\le\textbf{}\textbf{}avevmode\hbox{\small1\kern-3.8pt\normalsize1}}

\def\half{\frac{1}{2}}
\def\opone{\le\textbf{}\textbf{}avevmode\hbox{\small1\kern-3.8pt\normalsize1}}

\def\lbc{\left[}    \def\rbc{\right]}

\newcommand{\be}[1]{     \begin{eqnarray} \mbox{$\label{#1}$}   }

\newcommand{\ee}{\end{eqnarray}}

\begin{document}


\title[Entanglement Scaling of FQH states through
  Geometric Deformations]{Entanglement Scaling of Fractional Quantum Hall states through
  Geometric Deformations }

\author{Andreas~M.~L\"auchli, Emil J. Bergholtz, and Masudul Haque}
\address{Max-Planck-Institut f\"{u}r Physik komplexer Systeme,
N\"{o}thnitzer Stra{\ss}e 38, D-01187 Dresden, Germany}

\eads{\mailto{aml@pks.mpg.de}, \mailto{ejb@pks.mpg.de} and
\mailto{haque@pks.mpg.de}}

\begin{abstract} 
We present a new approach to obtaining the scaling behavior of the
entanglement entropy in fractional quantum Hall states from finite-size
wavefunctions.  By employing the torus geometry and the fact that the torus
aspect ratio can be readily varied, we can extract the entanglement entropy of
a spatial block as a continuous function of the block boundary length.  This
approach allows us to extract the topological entanglement entropy with an
accuracy superior to what is possible on the spherical or disc geometry, where
no natural continuously variable parameter is available.

Other than the topological information, the study of entanglement scaling is
also useful as an indicator of the difficulty posed by fractional quantum Hall
states for various numerical techniques.

\end{abstract}

\date{\today}


\maketitle

\section{Introduction}

Describing condensed matter phases using entanglement quantifiers from quantum
information theory is a rapidly growing interdisciplinary topic
\cite{AmicoFazioOsterlohVedral_RMP08, EisertRev}.  Of special interest are the
relatively rare cases where entanglement can provide information not readily
captured by conventional quantities such as correlation functions. This is the
situation for systems possessing \emph{topological order}
\cite{top-order_various}, where entanglement considerations have proven useful
\cite{KitaevPreskill_PRL06, LevinWen_PRL06, LiHaldane_PRL08}.  In particular
this has led to insight into the structure of fractional quantum Hall (FQH)
states \cite{HaqueZozulyaSchoutens, HaqueZozulyaSchoutens2, LiHaldane_PRL08,
  Fradkin_top-entropy-inChernSimons_JHEP08, ZozulyaHaqueRegnault,
  RegnaultBernevigHaldane_PRL09,MorrisFeder,Lauchli,ronny}.  In this Article
we focus on entanglement in this most realistic class of topologically ordered
states.
%

A prominent measure of entanglement is the von Neumann entropy of
entanglement, $S_A$, measuring the entanglement between a block ($A$) and the
rest ($B$) of a many-particle system in a pure state.  The entanglement
entropy $S_A = -\tr\lbc\rho_A\ln\rho_A\rbc$ is defined in terms of the reduced
density matrix, $\rho_A = \tr_B\rho$, obtained by tracing out $B$ degrees of
freedom from the system density matrix $\rho=|\psi\rangle\langle\psi|$, with
$|\psi\rangle$ denoting a ground state wave function.  In one dimension the
scaling behavior of the block entanglement entropy is well understood, see,
{\it e.g.}, Ref.~\cite{EisertRev}.  In two dimensions (2D), no such generic
classification exists.  However, for topologically ordered states in two
dimensions, the entanglement entropy contains topological information about
the state: $S_A$ scales as
\begin{equation}\label{eq_sa}
S_A =
\alpha{L} - n \gamma +\mathcal{O}(1/L),
\end{equation} 
where $L$ is the block boundary length, $\gamma$ characterizes the
topological field theory describing the state \cite{KitaevPreskill_PRL06,LevinWen_PRL06}, 
while $n$ counts the number of disconnected components of the boundary.
The value of $\gamma$ is related to the ``quantum
dimensions'' of the quasiparticle types of the theory, $\gamma =
\ln\mathcal{D}$, where $\mathcal{D}$ is the total quantum dimension.  For
Laughlin states at filling $\nu=1/m$, $\gamma = \half\ln{m}$.  For more
intricate FQH states, some examples of $\gamma$ values are   provided in
Refs.\ \cite{KitaevPreskill_PRL06, FendleyFisherNayak_JStatPhys07,
  HaqueZozulyaSchoutens2}.  
If $\gamma$ can be determined accurately, its value can in principle be used
to determine whether a topological phase belongs to the universality class of
a given topological field theory.

A numerical determination of $\gamma$ and $\alpha$ requires information about
$S_A$ for a number of different boundary lengths, $L$.  In
Ref.~\cite{HaqueZozulyaSchoutens} such information was obtained from
finite-size FQH wavefunctions by approximating the spatial partitioning by
partitioning of the discrete set of Landau level orbitals.
Refs.~\cite{HaqueZozulyaSchoutens,HaqueZozulyaSchoutens2,ZozulyaHaqueRegnault}
used spherical geometries, and explored ways of extrapolating entanglement
information from such geometries to the thermodynamic limit.
Ref.~\cite{MorrisFeder} used disk geometries to calculate $\gamma$ for bosonic
FQH wavefunctions.  The accuracy in the determination of $\gamma$ from
finite-size wavefunctions on these geometries remains disappointing
(10\%--30\% for the simplest Laughlin states).  Improved methods for
calculating $\gamma$ are thus sorely needed.

In this work, we report a significant advance in this direction, through the
use of the torus geometry and the fact that the aspect ratio (circumference)
of the torus can be varied continuously without drastically altering the torus
setup or symmetry.  Varying the circumference changes the length of the
boundary between $A$ and $B$.  No natural analogous continuous parameter
exists in the other geometries, so that in those cases each system size and
bipartition provides only one point in parameter space.  Exploiting the
continuous parameter, we present a procedure that leads to an accuracy in
$\gamma$ down to a few percent.  Our analysis also provides a visual and
physical indication of the reliability of the extracted $\gamma$ value.

Previously, Ref.~\cite{friedman} reported a na\"ive modification of the sphere
algorithm of Ref.~\cite{HaqueZozulyaSchoutens} to the torus geometry with
fixed aspect ratio.  We will show why this analysis was inappropriate and
based on an extrapolation procedure that is not meaningful for the torus
geometry.

In addition to the topological content of the subleading term in 
(\ref{eq_sa}), the dominant linear term itself is also of some importance.  The
rate of entanglement growth, $\alpha$, indicates how challenging the state is
to simulate on a classical computer, through a one-dimensional algorithm like
DMRG \cite{White,dmrgrev}, or through recently-proposed true two-dimensional
algorithms like PEPS \cite{PEPS} or MERA \cite{MERA}.  DMRG has been used to
simulate FQH states 
\cite{Shibata,bk03,feiguin,kovrizhin}, and
these states pose a future challenge for two-dimensional algorithms currently
under development.

The calculation of the topological entanglement entropy $\gamma$ is of
significant current interest, not only for FQH states but also for various
other topologically ordered states.  For the zero-temperature Kitaev model, it
is relatively easy to calculate $\gamma$; so the concept has been used in
exploring issues such as temperature effects and quantum phase transitions
\cite{CastelnovoChamon_PRB07, CastelnovoChamon_topol-QPT_PRB08,
  Hamma-etal_KitaevQPT_PRB08, IblisdirPerezGarciaAguadoPachos_PRB09}.  For
quantum dimer models and related states, considerations of entanglement
scaling are more intricate
\cite{FurukawaMisguich_PRB07,PapanikolaouRamanFradkin_PRB07}, of difficulty
comparable to FQH states.  
More generally, entanglement scaling in 2D states of all kinds has become the
focus of intense study at present \cite{entanglement_2Dscaling, EisertRev,integer}.

In Section \ref{setup} we show how the torus geometry allows us to map the
interacting Landau level (LL) problem onto a one-dimensional lattice model,
appropriate for studying bipartite entanglement.  In Section
\ref{sec_lA-dependence_degeneracy} we outline the general behavior of
entanglement entropy on the torus geometry and deal with the issue of torus
degeneracies.  In Section \ref{results} we present our main results, including
an analysis leading to the determination of $\gamma$.  The concluding Section
\ref{discussion} connects to the existing literature and discusses
implications of our results.

\section{Torus setup -- geometry, partitioning}\label{setup}

\begin{figure}[ht]
\centerline{
\includegraphics*[width=0.9\textwidth]{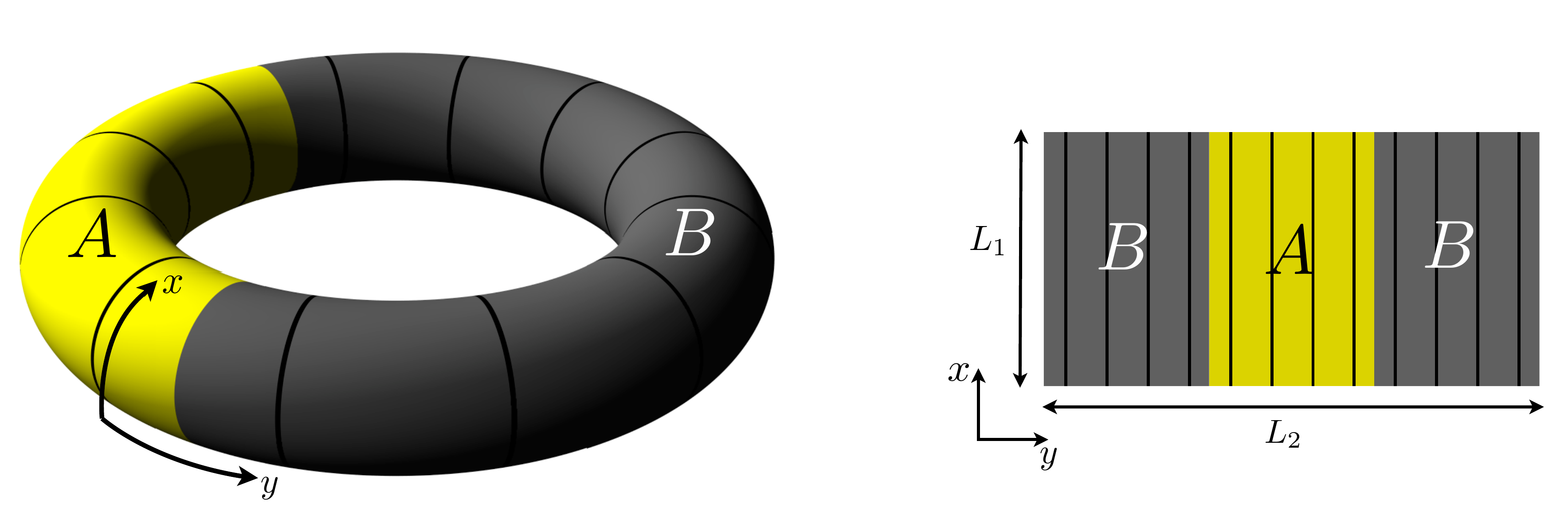}}
\caption{  \label{fig:torus_geometry}
{\bf Geometry of the torus and bipartitioning}.  
The lowest Landau level is spanned by orbitals which in the Landau gauge are
centered along the circles shown.  On the right, we represent the torus as a
rectangular region with periodic boundary conditions in both directions.  The
dimensions of this rectangle ($L_1$, $L_2$) are the two circumferences of the
torus.  The example shown here has $N_s=12$ orbitals with $l_A=4$ orbitals in
the $A$ block.  }
\end{figure}

We study an $N$-electron system on a torus (see Fig.~\ref{fig:torus_geometry})
with periods $L_1, L_2$ in the $x$- and $y$-directions, satisfying
$L_1L_2=2\pi N_s$ in units of the magnetic length.  The integer $N_s=N/\nu$ is
the number of magnetic flux quanta.
In the Landau gauge, $\mathbf{A}=By\mathbf{\hat{x}}$, a
basis of single particle states in the lowest Landau level can be taken as
\begin{equation}\psi_{j}=\pi^{-1/4}L_1^{-1/2}\sum_{m} e^{i(\frac{2\pi}{L_1}j+mL_2)x}
e^{-(y+\frac{2\pi}{L_1}j+mL_2)^2/2}\label{psi}\end{equation}
with $j=0,1,...,N_s-1$. The states
$\psi_j$ are 
centered along the lines $y=-2\pi j/L_1$.  Thus the
$y$-position is given by the $x$-momentum $j$.

A generic translation-invariant two-body interaction Hamiltonian, acting within a Landau level, can be written as
\begin{equation}
\label{ham}
H =\sum_n \sum_{k > |m|} V_{km}c^\dagger_{n+m}c^\dagger_{n+k}c_{n+m+k}c_n \ \ ,
\end{equation}
where $c^\dagger_m$ creates an electron in the state $\psi_m$ and $V_{km}$ is
the amplitude for two particles to hop symmetrically from separation $k+m$ to
$k-m$ \cite{bk}.  Hence, the problem of interacting electrons in a Landau
level maps onto a one-dimensional, center-of-mass conserving, lattice model
with lattice constant $2\pi/ L_1$. This provides a natural setting for
defining entanglement, by bipartitioning the system into blocks $A$ and $B$,
which consist respectively of $l_A$ consecutive orbitals and the remaining $l_B=N_s-l_A$
orbitals (Fig.~\ref{fig:torus_geometry}).  Since the orbitals
are localized in the direction of the lattice, this is a reasonable
approximation to spatial partitioning, as on the sphere
\cite{HaqueZozulyaSchoutens,HaqueZozulyaSchoutens2,LiHaldane_PRL08,ZozulyaHaqueRegnault,RegnaultBernevigHaldane_PRL09}.

Because this partitioning implies two disjoint edges between the blocks, each
of length $L_1$, the entanglement entropy should satisfy the following
specific form of (\ref{eq_sa}):
\begin{equation}\label{satorus}
S_A(L_1) =2\alpha{L_1} -2\gamma +\mathcal{O}(1/L_1).
\end{equation} 
Thus our setup should yield a linear scaling form of the entropy with the $L_1=0$
intercept at $-2\gamma$.

In this work, we obtain ground states of (\ref{ham}), in the orbital basis
(\ref{psi}) using the Lanczos algorithm for numerical diagonalization.  We
study bipartite entanglement in these ground states.  Apart from diagonalizing
the Coulomb problem we also consider pseudopotential interactions
\cite{haldane83,Trugman-K} which have the Laughlin states \cite{laughlin83} as
exact ground states. The largest Hilbert space sizes considered are 208'267'320 for 39 orbitals
at $\nu=1/3$, 19'692'535 for 45 orbitals at $\nu=1/5$ and 66'284'555 for 35 orbitals at $\nu=2/5$. The simulations
are however currently limited by the size of the reduced density matrices to be calculated and fully diagonalized.



Fractional quantum Hall states have degenerate ground states on the torus
geometry.  It is convenient to label the ground states by their corresponding
thin torus (or Tao-Thouless, TT) patterns \cite{bk,seidel,anderson,tt,RH94}.
For example, for $\nu=1/3$ there are three degenerate states, which correspond
to the TT configurations
\begin{eqnarray}
100100100\Big{|}100100100100100100\Big{|}100100100\nonumber\\
010010010\Big{|}010010010010010010\Big{|}010010010\nonumber\\
001001001\Big{|}\!\underbrace{001001001001001001}_{A}\!\Big{|}001001001 \ ,\label{ttonethird}
\end{eqnarray}
for $N_s=36$.  Here the positions of $1$'s indicate the positions (or,
equivalently, the transverse momenta) of filled single particle states.  An
equal partitioning ($l_A=l_B=N_s/2$) is illustrated.  

In general, abelian FQH states at $\nu=p/q$ have $q$ degenerate ground states,
related to each other through translation and corresponding to $q$ thin-torus
patterns, each composed of unit cells with $p$ electrons on $q$ sites.  These
states are ground states for generic (two-body) interactions as
$L_1\rightarrow 0$ \cite{bk}.  For non-abelian states there is an enhanced
degeneracy and the corresponding thin torus patterns are not simply
translations of each other.

The thin torus states are unentangled product states, in the orbital basis. As
$L_1$ is increased from zero, fluctuations on top of the TT states will make
the states entangled. A crucial property of the FQH states is that their bulk
versions are, for appropriate interactions, adiabatically connected to their
respective TT states and the gap is finite for all $L_1$
\cite{bk,seidel}. This allows us to probe the response of these states as the
geometry is deformed.  Such deformations have also been considered to extract
properties such as the Hall viscosity \cite{read09,haldane09}, to put
consistency conditions of FQH states \cite{seidel10}, to find instabilities to
competing states (see {\it eg} \cite{cdw,Yang}) and to deform the torus to the
solvable thin torus limit as discussed above.  All the three fractions studied
in this paper ($\nu=1/3$, $1/5$, and $2/5$) are, for the pseudopotential as
well as the Coulomb interactions, continuously connected to their TT states.

\begin{figure}[ht]
\centerline{
\includegraphics*[width=0.9\linewidth]{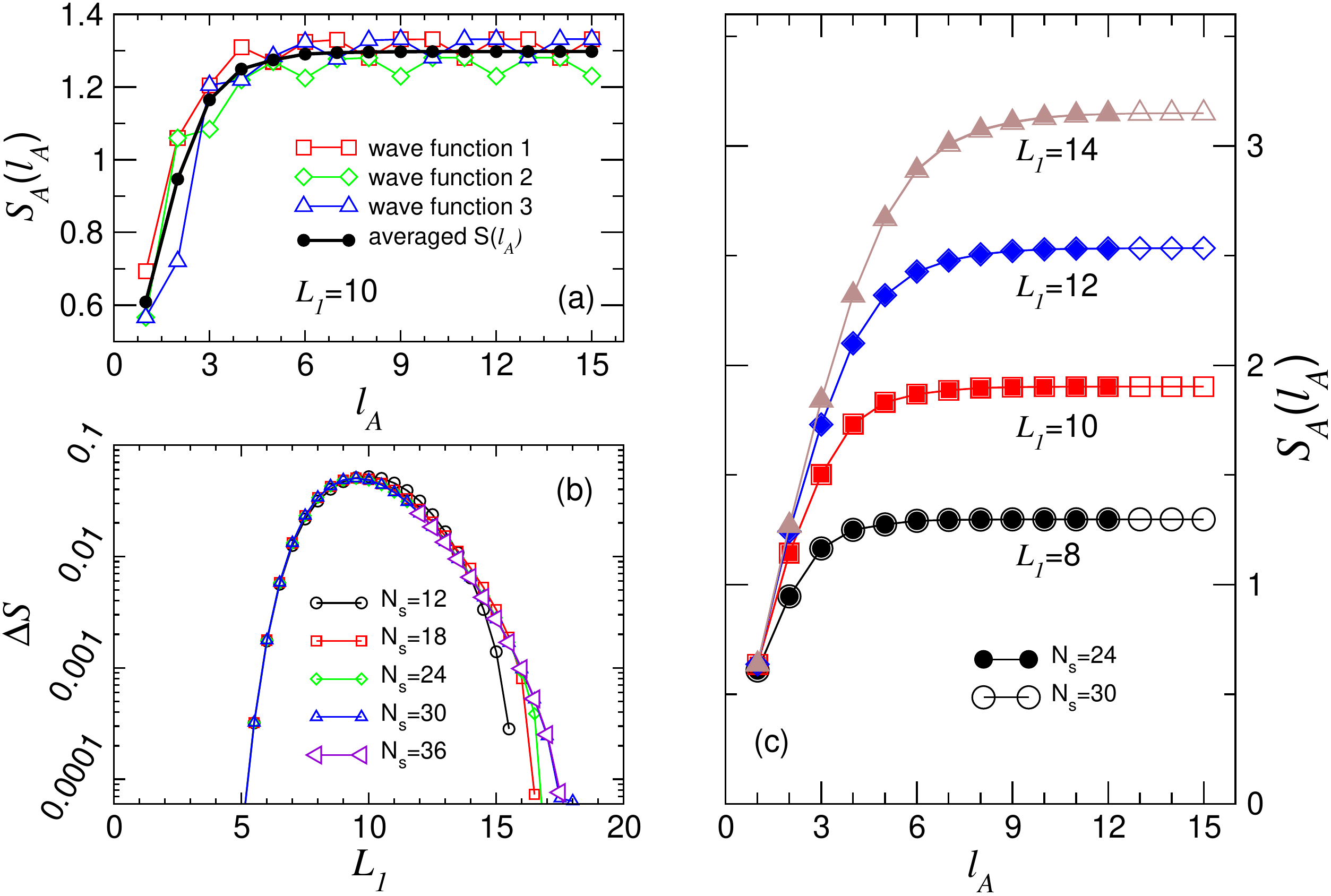}
}
\caption{ {\bf Degeneracy averaging and $l_A$ dependence}.  (a)
  Entanglement entropy in different degenerate sectors for the $\nu=1/3$
  Laughlin state, and their arithmetic average.  (b) difference between
  degeneracy-averaged $S_A$ and largest individual-sector $S_A$, as a function
  of $L_1$.  They differ significantly only for intermediate $L_1$. 
  (c) Degeneracy-averaged  $S_A$ versus $l_A$, for different $L_1$
  values. 
\label{fig:averaged}}
\end{figure}

\section{Degeneracy averaging, Area law at constant $L_1$} \label{sec_lA-dependence_degeneracy}

For any finite $L_1$ the charge density modulations of the TT pattern will
prevail to some extent, leading to different entanglement in the $q$
degenerate ground states.
We illustrate this in Fig.~\ref{fig:averaged}(a) where we plot the
entanglement entropy $S_A(l_A)$ as a function of $l_A$ in the three degenerate
$\nu=1/3$ Laughlin wave functions. For each $l_A$ two out of the three
entanglement entropies are equal, while the third one is different, as can be
inferred from examining the partitionings shown in \Eref{ttonethird}. 
Two of the TT patterns have 1-0 and 0-1 cuts at the block boundaries,
  while the third has only 0-0 cuts.  Since the microscopic environment at the
  two boundaries determines the entanglement spectrum \cite{Lauchli} and hence
  the entanglement entropy, this implies that two of the entanglement
  entropies are equal at each $l_A$.

The $S_A(l_A)$ each have prominent oscillatory behavior.  However, we find
that the \emph{arithmetic mean} of the three individual entanglement entropies
is remarkably free of oscillations.  We will thus base all our following
discussions on the degeneracy-averaged entropy
\footnote{ In principle the averaging could be done in other ways, for
  example, one could average over the density matrices $\rho\rightarrow
  \sum_{i=1}^{N_{GS}}\rho^{(i)}/N_{GS}$ or reduced density matrices, and then
  compute the entanglement entropy from this averaged matrix.  We do not
  however pursue such alternate averaging procedures in the present work.}.
Ultimately this averaging will become unimportant for very large $L_1$, as
Fig.~\ref{fig:averaged}(b) shows that the difference $\Delta S$ between the
maximum and the mean entanglement entropies in the $q$ sectors vanishes
rapidly at large $L_1$ (starting to decrease after $L_1 \sim 10$ in the
Laughlin $\nu=1/3$ case shown).

In Fig.~\ref{fig:averaged}(c) we study the $l_A$ and $N_s$ dependence for a
given $L_1$. At constant $L_1$ we expect the entropy to saturate once $l_A$ is
large enough, since the block boundary length ($2 L_1$) is held constant.
This is indeed what is found numerically in Fig.~\ref{fig:averaged}(c). The
length scale controlling the saturation is the (real space) correlation length
$\xi_r$ in the $y$ direction of the incompressible FQH liquid. 
The correlation length $\xi_o$ measured in number of orbitals is expected to
scale as $\xi_o \sim \xi_r \times L_1/2\pi$.
The saturation of the entanglement entropy for large
$l_A$ is in complete analogy to the area law for one-dimensional gapped
systems~\cite{VidalLatorreRicoKitaev_PRL03, hastings07}.  It is this
saturation value $S_A(L_1)$ of the entanglement entropy obtained for $l_A\gg
\xi_o$ that will be analyzed in the following.  To avoid the finite size
effect as far as possible we consider $l_A=N_s/2$ for $N_s$ even, and $l_A=(N_s-1)/2$ for
$N_s$ odd, in the rest of this
article.  From Fig. \ref{fig:averaged}(c) we can again infer that the averaged
$S_A$ indeed has a much smoother dependence on the block size compared to the
$S_A$ for the individual degenerate states.


\section{Accessing the scaling regime} \label{results}

\begin{figure}[t]
\centerline{
\includegraphics*[width=0.8\linewidth]{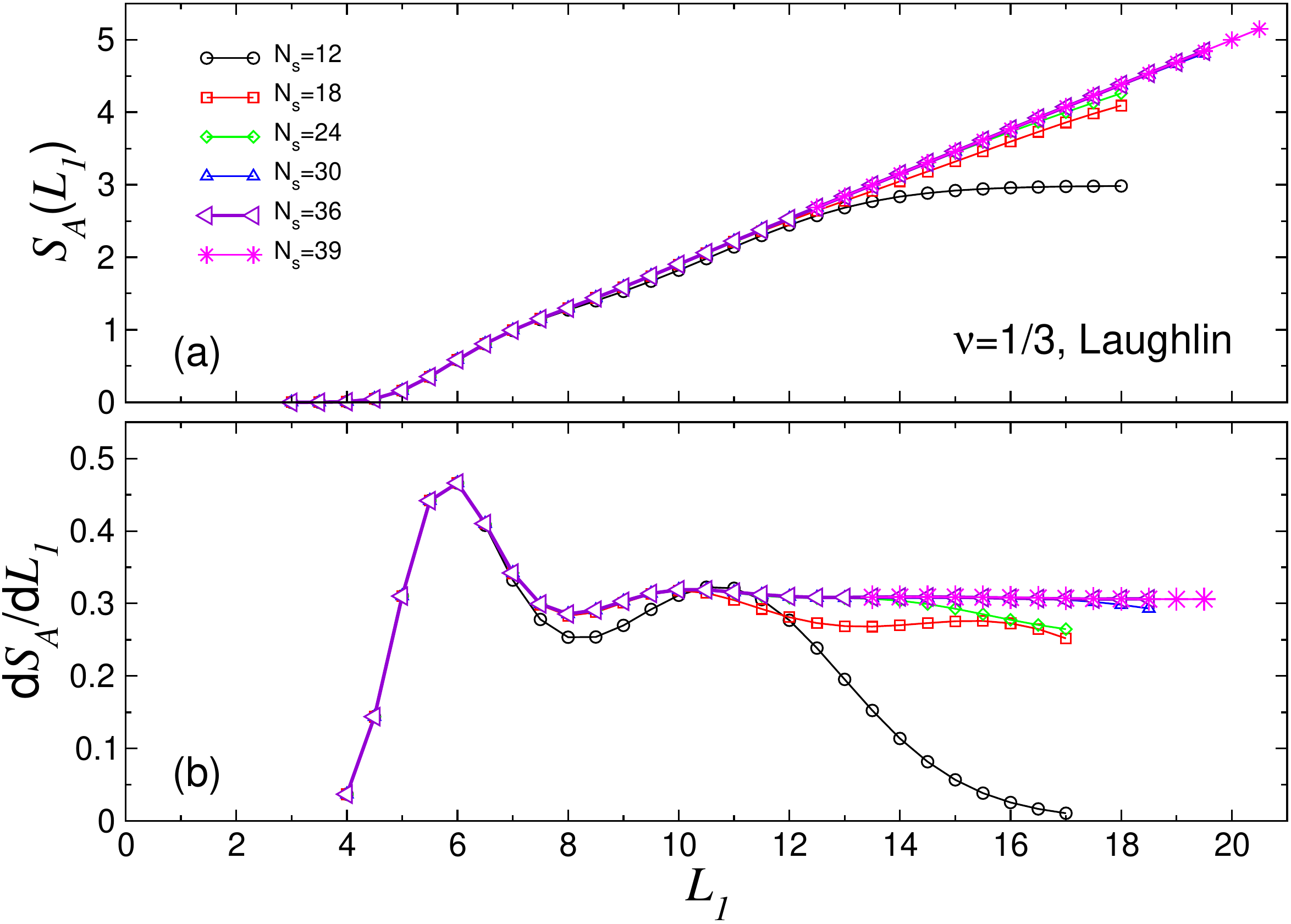}
}
\caption{
{\bf $\nu=1/3$ Laughlin state: Entanglement entropy}, (a) $S_A$ and (b) its derivative $\mathrm{d}S_A/\mathrm{d}L_1$ for the Laughlin state at $\nu=1/3$
as a function of $L_1$. From the plateau behavior in $\mathrm{d}S_A/\mathrm{d}L_1$ for $L_1\gtrsim 12$ we infer $\alpha \approx 0.153(2)$.
\label{fig:ee_onethird_laughlin}
}
\end{figure}

In this Section we provide our main numerical results and discuss how $\alpha$
and $\gamma$ can be extracted by continuously varying $L_1$.
As mentioned above, from now on $S_A$ refers to the equal-partitioning
entanglement entropy ($l_A=N_s/2$ for $N_s$ even, and $l_A=(N_s-1)/2$ for
$N_s$ odd).

\paragraph{$\nu=1/3$  Laughlin state ---}

Fig.~\ref{fig:ee_onethird_laughlin} shows the behavior of $S_A(L_1)$ (a) and
its derivative $dS_A/dL_1$ (b), for the Laughlin state at fraction $\nu=1/3$,
arguably the most prominent and also the simplest FQH state.  In this figure
and subsequently, we use a five-point formula to numerically obtain
derivatives.  The entanglement entropy $S_A(L_1)$ remains minuscule until
$L_1\sim3$, and then gradually changes to the expected linear increase
behavior which is reached around $L_1\sim 7$.  There are oscillations on top
of the linear behavior, which are more prominent in the derivative plot (b).
The oscillations can be interpreted as an interplay between the finite
circumference along the $x$ direction (finite $L_1$) and the interparticle
distance. The oscillations die off as a function of $L_1$, so that if $N_s$ is
large enough one can get the scaling form at large $L_1$.

For small $L_1$ the finite size convergence is essentially perfect. At larger
$L_1$, the $S_A(L_1)$ and $dS_A/dL_1$ curves show stronger dependence on
$N_s$.  The $N_s$-dependence shows up first for the smallest system sizes and
at increasing $L_1$ for progressively larger system sizes.  This reflects the
fact that, for any finite-size system, at very large $L_1$ the edges of $A$
get too close (small $L_2$) and cannot be thought of as independent
\cite{Lauchli}.  In particular, once $L_1$ exceeds some value we enter the
``dual thin torus'' or ``thick torus'' limit \cite{RH94,fat,seidel10}, and the
entanglement entropy levels off to some saturation value.  Corresponding to
the saturation of $S_A(L_1)$, the derivative $dS_A/dL_1$ drops off to zero
after some $L_1$.

Thus, the scaling form of (\ref{satorus}) is valid only in a window of $L_1$,
after the oscillations have subsided but before $S_A(L_1)$ saturates, or shows
other precursor finite size effects.  This plateau region can be seen clearly
in the $dS_A/dL_1$ curve for the $\nu=1/3$ Laughlin data for $L_1 \gtrsim
12$. The finite size convergence of the data also provides a clear signal
showing whether the bulk scaling regime is reached or whether (geometrical)
finite size effects are still significant in the numerically accessible $L_1$
regime.

\begin{figure}[t]
\centerline{
\includegraphics*[width=0.8\linewidth]{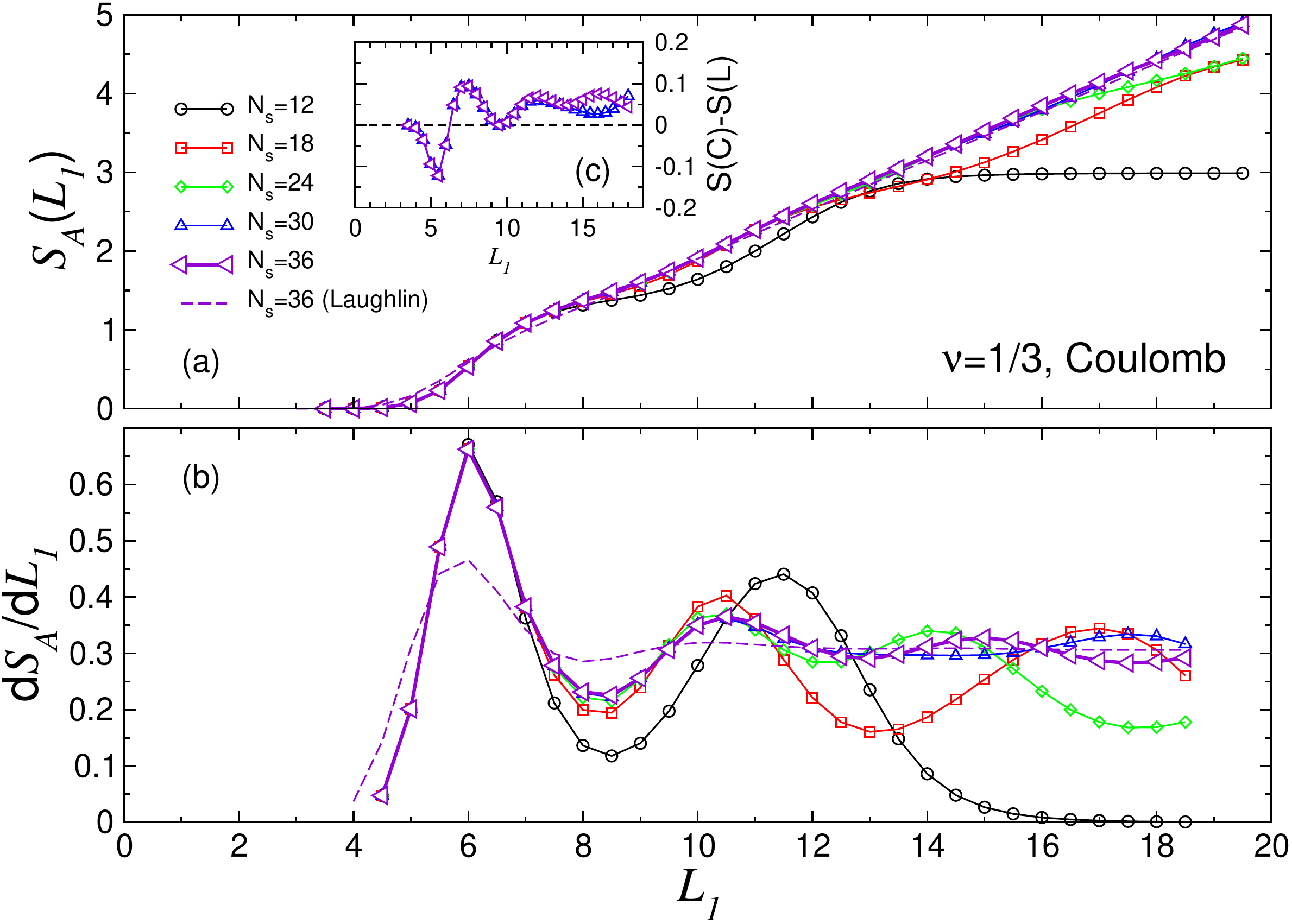}
}
\caption{ {\bf $\nu=1/3$ Coulomb ground state: Entanglement entropy}, (a)
  $S_A$ and (b) its derivative $\mathrm{d}S_A/\mathrm{d}L_1$ for the Coulomb
  ground state at $\nu=1/3$ as a function of $L_1$.  One curve for the
  Laughlin state is also shown for comparison (dashed line).
In the inset (c) the difference in $S_A(L_1)$ between the Coulomb and the
Laughlin is displayed (for $N_s=30, 36$).
\label{fig:ee_onethird_coulomb}
}
\end{figure}


\paragraph{$\nu=1/3$  Coulomb ground state ---}

Fig.\ \ref{fig:ee_onethird_coulomb} plots (a) $S_A(L_1)$ and (b) $dS_A/dL_1$
for the Coulomb ground states at $\nu=1/3$.  While this state has somewhat
more severe finite-size effects and oscillatory behaviors compared to the
model Laughlin state of Fig.~\ref{fig:ee_onethird_laughlin}, we note that the
scaling form of their entanglement entropies are very similar.  To further
highlight this fact we plot the difference between the entanglement entropies
as a function of $L_1$ in the inset of Fig.~\ref{fig:ee_onethird_coulomb}(c).
The similarity of entanglement entropies of the two states is not unexpected
from the perspective that the states have a large overlap for all $L_1$
\cite{Yang}, but it is nevertheless interesting considering that a more
"generic" state such as the Coulomb state is expected to have larger
entanglement.
One could thus have expected the Coulomb state to have a larger $\alpha$, as
defined in Eq.\ \ref{eq_sa}, but Fig.~\ref{fig:ee_onethird_coulomb}(b)
suggests a very similar $\alpha \approx 0.15(1)$.

\begin{figure}[t]
\centerline{
\includegraphics*[width=1\linewidth]{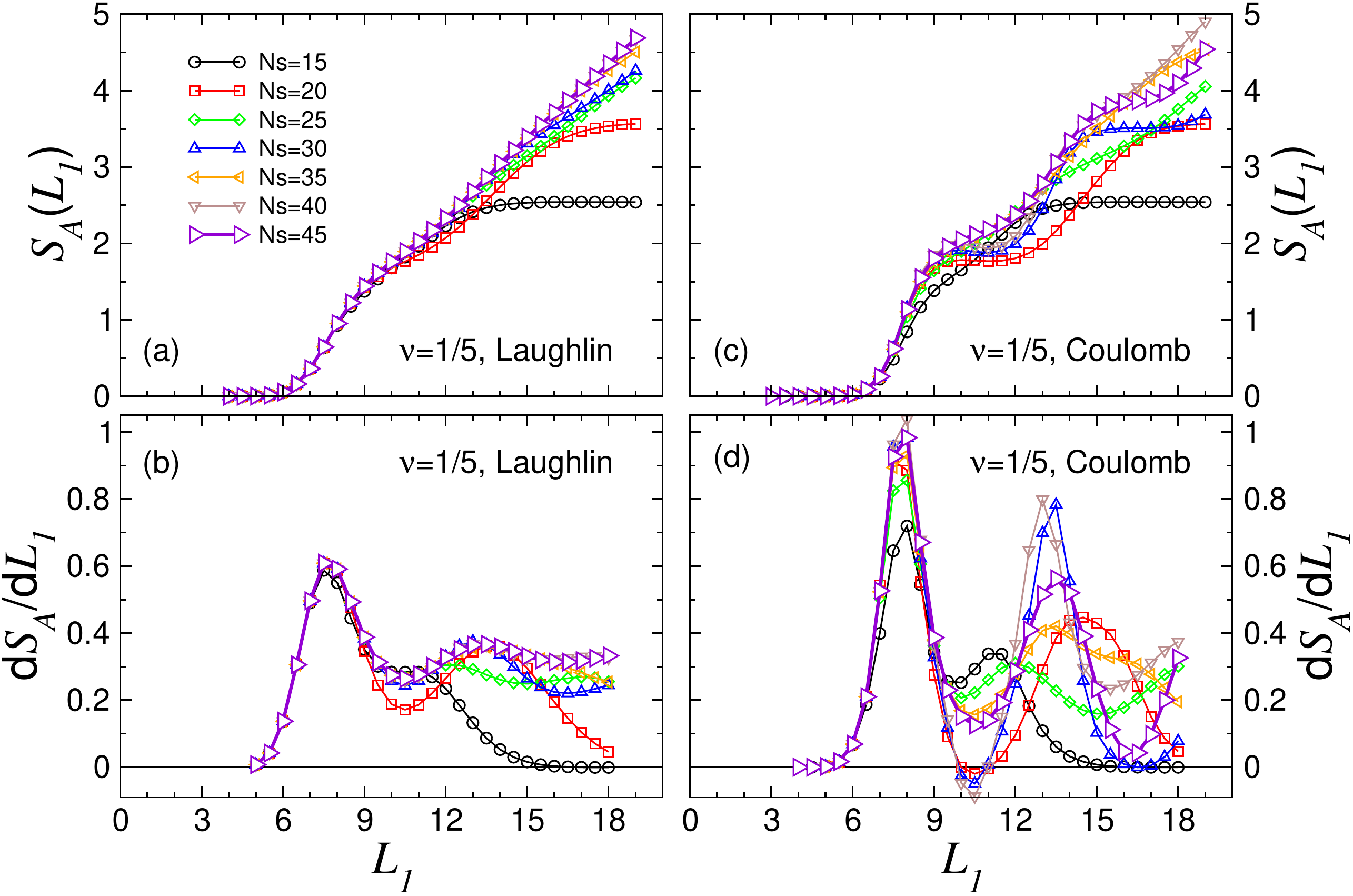}
}
\caption{
{\bf $\nu=1/5$ Laughlin vs Coulomb:} 
Entanglement entropy (a) and its derivative (b) for the $\nu=1/5$ Laughlin
wave functions. Entanglement entropy (c) and its derivative (d) for the $\nu=1/5$ 
Coulomb ground state. 
\label{fig:S_dSdL_one_fifth}
}
\end{figure}

\paragraph{$\nu=1/5$ ---}
Fig.~\ref{fig:S_dSdL_one_fifth} shows the $S_A(L_1)$ and $dS_A/dL_1$ behaviors
at $\nu=1/5$, for both the Laughlin (a) \& (b) and the Coulomb ground state
(c) \& (d).  As expected, the finite-size oscillations are much more severe in
these states.  This is expected as the interparticle distance is larger; thus
larger systems should be required to reach the scaling regime.  Moreover, the
proximity to the Wigner crystal phase makes the Coulomb ground state deviate
more substantially from the Laughlin state than is the case at $\nu=1/3$
\cite{lam,Yang}.  While we are able to get an almost $N_s$-converged $S(L_1)$
curve up to $L_1\sim 18$ for the Laughlin state [leading to a rough estimate
  of $\alpha\approx 0.17(2)$], the finite size effects in the Coulomb ground
state are so severe that no meaningful extraction of $\alpha$ is possible with
current system sizes.

\begin{figure}[t]
\centerline{
\includegraphics*[width=0.9\linewidth]{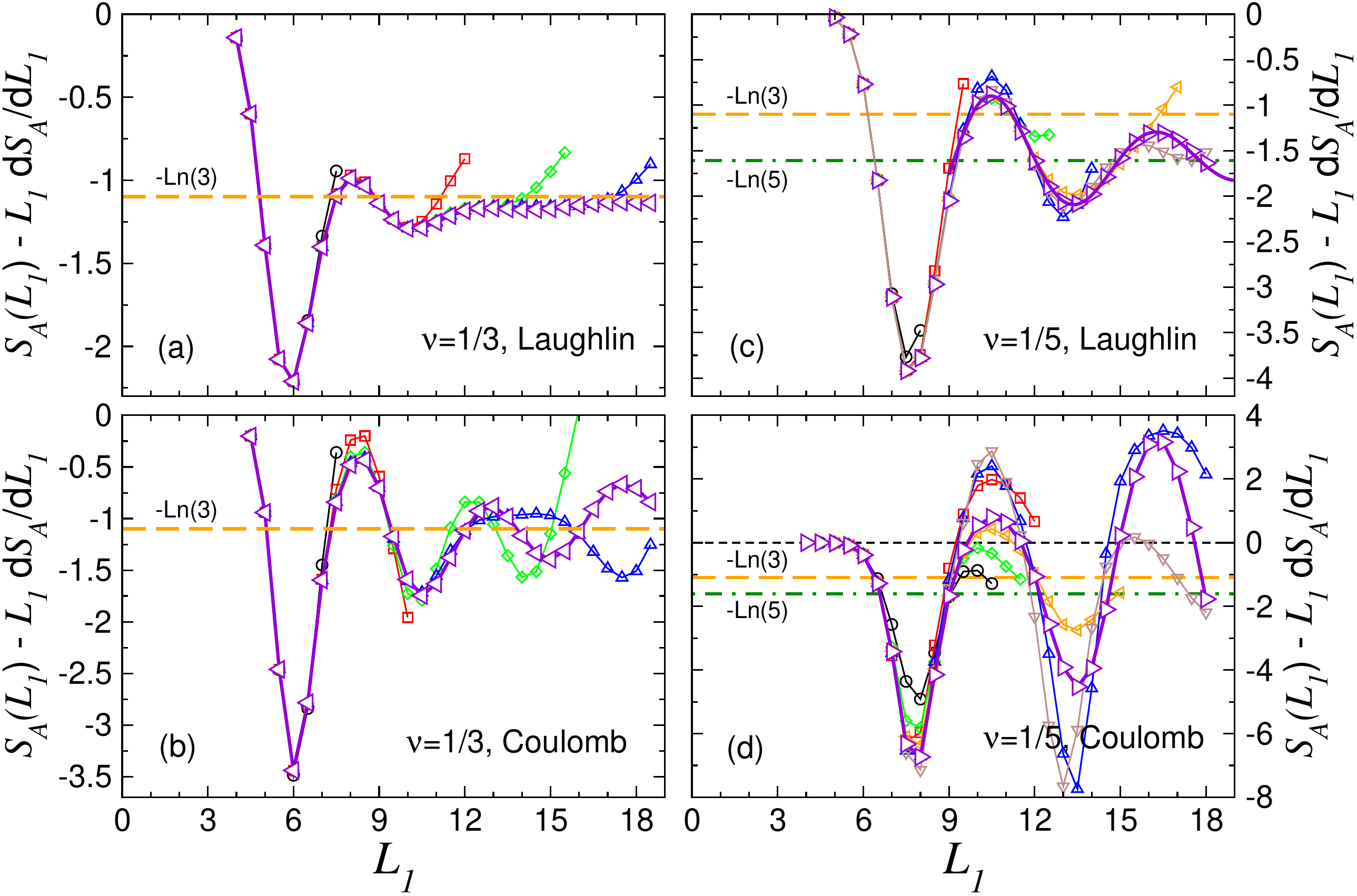}
}
\caption{ {\bf $L_1$-local extraction of $\gamma$}.  The intercept of local
  linear approximations to the $S_A(L_A)$ curves, \emph{i.e.}, $S_A(L_1)-L_1\times
  dS/dL_1$, plotted as a function of $L_1$.  In the scaling regime, this
  quantity should give $-2\gamma$.  The symbols for $\nu=1/3$ ($\nu=1/5$) are
  the same as in
  Fig.~\protect{\ref{fig:ee_onethird_coulomb}}~(Fig.~\protect{\ref{fig:S_dSdL_one_fifth}}).
  Theoretically expected $-2\gamma$ values are shown as dashed horizontal
  lines. In panel (c) the solid line through the largest size data is the fit obtained using \Eref{eqn:damped_fit}. 
\label{fig:gamma_panel}
}
\end{figure}

\paragraph{Extraction of the topological entanglement entropy $\gamma$ --- }
In Fig.~\ref{fig:gamma_panel} we show calculations of $\gamma$ for the Laughlin
state at $\nu=1/3$ (a) and $\nu=1/5$ (c) as well as for the Coulomb ground state at the same
fractions (b) \& (d).
Evaluating the $L_1$ derivative using a centered 5-point formula, we plot
$S_A(L_1)-L_1\times dS/dL_1$ as a function of $L_1$.  This quantity is the
intercept of a linear approximation made to the $S_A(L_1)$ curve locally at
each $L_1$.  It should take the value $-2\gamma$ in the scaling region, see
\Eref{satorus}.  Not surprisingly, the intercept oscillates at intermediate
$L_1$, has a plateau in the ``scaling window'' described above, and then moves
off to a large positive value when $L_1$ is yet larger entering the thick
torus regime.  The plateau region value gives us the best estimate for the
topological entanglement entropy. A significant advantage of our analysis is
that, by examining the $dS_A/dL_1$ curve (and its $N_s$ dependence), we can
identify the correct window of $L_1$ values to use.

In Fig.~\ref{fig:gamma_panel}(a) the $\nu=1/3$ Laughlin shows such a clear
plateau region.  The plateau region value around $L_1 \sim 18$ gives us the
best estimate for the topological entanglement entropy $\gamma\approx
0.565(5)$, to be compared to the theoretical expectation
$\gamma=\ln(3)/2\approx 0.5493$.  The difference amounts to only 3 percent in
this ideal case.

For the $\nu=1/5$ Laughlin [Fig.~\ref{fig:gamma_panel}(c)], the finite-size issues are significantly larger,
and the oscillations have not yet damped out at accessible sizes.
However, one can take the average of the oscillating values to get a
reasonable estimate of $\gamma$. We use a simple damped oscillation fitting 
ansatz of the form:
\begin{equation}
f(L_1) = -2\gamma + a \times \exp[-b L_1] \times \sin (c L_1- d),
\label{eqn:damped_fit}
\end{equation}
and fit the $N_s=45$ curve for $L_1 > 9$, yielding an estimate of
$\gamma\approx 0.81$. This value again compares very favorably to the
theoretical expectation $\gamma=\ln(5)/2\approx 0.8047$. 

At each of these fractions, the finite-size convergence is worse for the
Coulomb ground state compared to the Laughlin ground state. While for the
$\nu=1/3$ Coulomb [Fig.~\ref{fig:gamma_panel}(b)] a fitting analysis along the
lines of the $\nu=1/5$ Laughlin still provides a reasonable $\gamma$ estimate:
$\gamma \approx 0.60$, the $\nu=1/5$ Coulomb state
[Fig.~\ref{fig:gamma_panel}(d)] clearly does not allow a meaningful $\gamma$
extraction from system sizes presently reachable through numerical exact
diagonalization.

\begin{figure}[t]
\centerline{
\includegraphics*[width=0.8\linewidth]{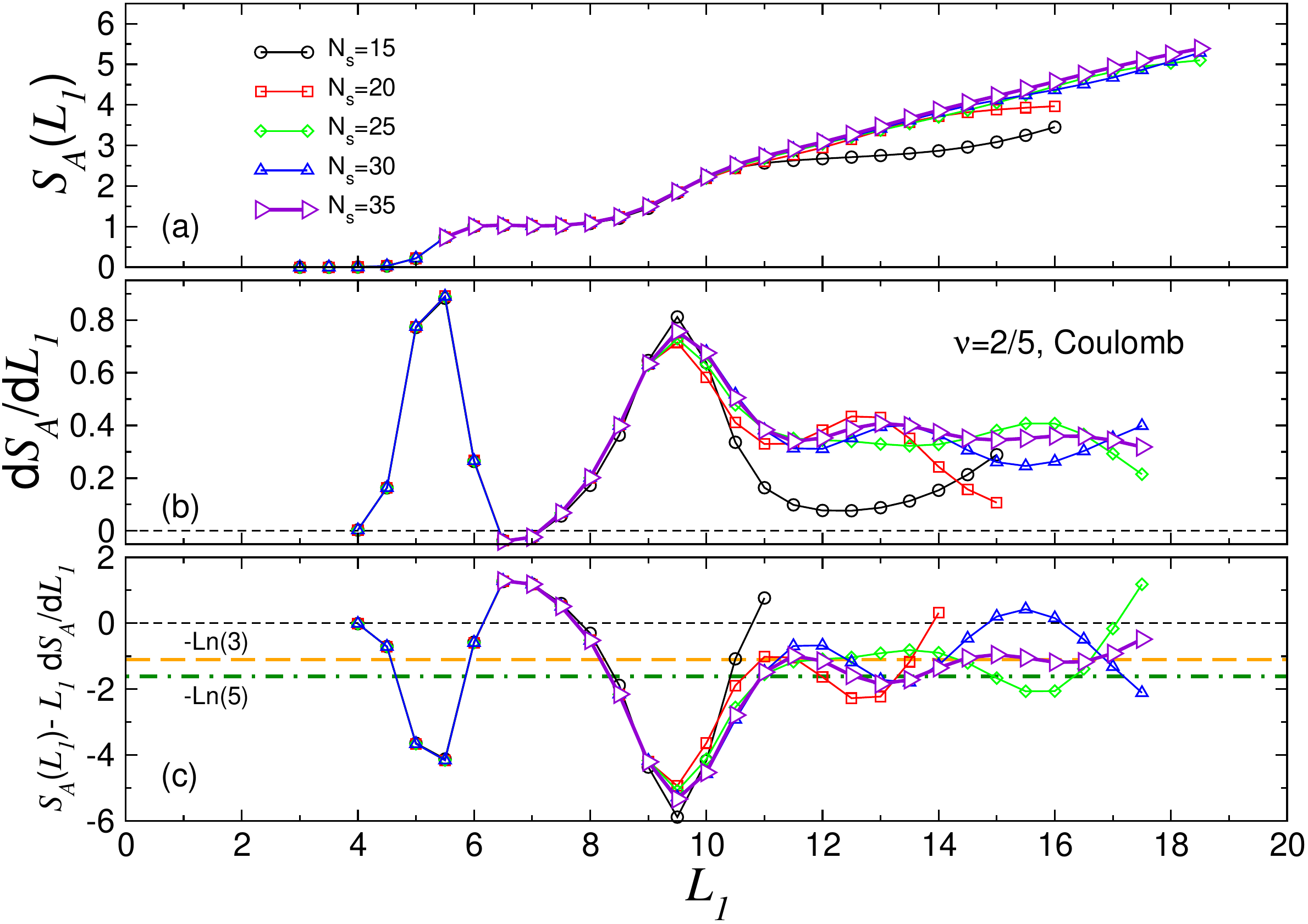}
}
\caption{
{\bf$\nu=2/5$ Coulomb ground state.}   
(a)
  $S_A$ and (b) its derivative $\mathrm{d}S_A/\mathrm{d}L_1$ for the Coulomb
  ground state at $\nu=2/5$ as a function of $L_1$.  (c) Estimation of
  $-2\gamma$ through a plot of the $L_1$-local intercept against  $L_1$.  
\label{fig:twofifth}
}
\end{figure}

\paragraph{$\nu=2/5$ Coulomb interaction ---}
Finally, in Figure \ref{fig:twofifth} we consider the Coulomb ground state at
filling $\nu=2/5$, {whose $\gamma$ value has not been studied numerically so
  far}.  This state is, for all $L_1$, well described by the torus version
\cite{torus} of the Jain \cite{jain89} (or, equivalently, the hierarchy
\cite{haldane83}) state.  The finite-size effects are somewhat less severe
than the $\nu=1/5$ Coulomb case [Figs.~\ref{fig:S_dSdL_one_fifth}(c,d) and
  \ref{fig:gamma_panel}(d)].  One obtains an entanglement growth rate of
$\alpha\approx 0.188(16)$.  While the $N_s$-convergence is not good enough for
a precise determination of $\gamma$ (expected to be $\half\ln5$), examination
of the largest two available sizes suggests that two or three additional sizes
may be enough to provide an estimate at the $\sim$10\% accuracy level.


\section{Discussion}\label{discussion}

In this article we have shown how continuous geometric deformations of the
torus can be employed to explore the scaling form of the entanglement entropy.
This has allowed us to propose a method for determining the topological part,
$\gamma$, from finite-size wavefunctions, to greater precision compared to
earlier analyses which did not utilize any continuous parameter.
Our analysis  indicates that current state-of-the-art system sizes are
enough to obtain reliable $\gamma$ calculations for the simplest fractional
quantum Hall states (Laughlin states), but that more intricate states would
require larger sizes than currently accessible, in order to reach the scaling
limit.  Our procedure provides a clear method for identifying whether the
scaling window has been reached or not.

\begin{figure}[ht]
\centerline{
\includegraphics*[width=0.9\linewidth]{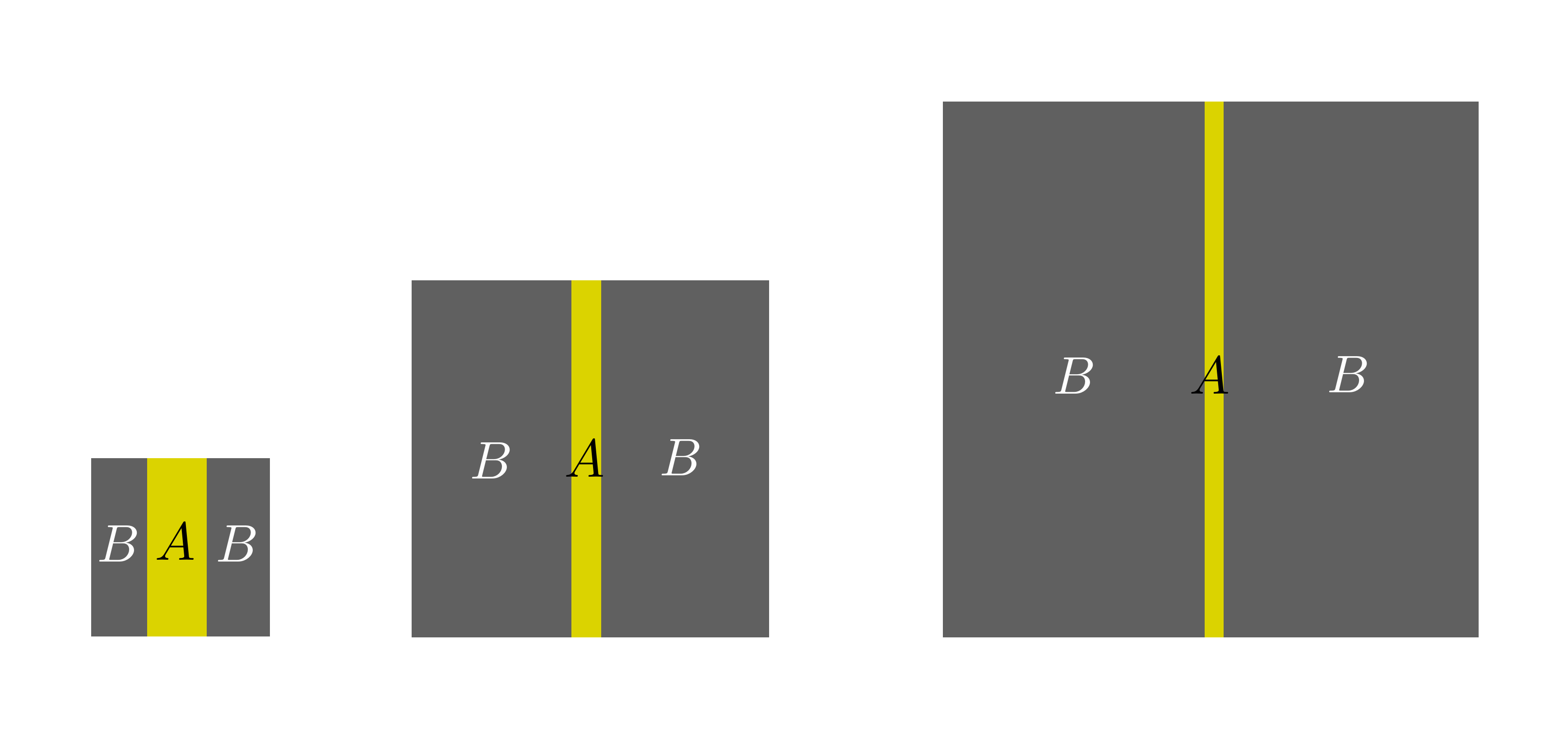}
}
\caption{ {\bf Extrapolation at fixed aspect ratio and fixed $l_A$}. Fixed
  $l_A$ implies that the area covered by the region $A$ is constant.  Such
  extrapolation does not lead to a well-defined limit as the limiting case is
  one with infinitesimal thickness and infinite boundary length.
\label{fig:FriedmanLevineScaling}
}
\end{figure}

There has been an earlier report of entanglement entropy and $\gamma$
calculations on the torus \cite{friedman}, using fixed aspect ratio,
$L_1/L_2=1$.  Ref.~\cite{friedman} performed $N_s\rightarrow\infty$
extrapolations at fixed $l_A$, and expected the extrapolated values to scale
as $c_1\sqrt{l_A}-2\gamma$.  We illustrate such a fixed $l_A$ extrapolation in
Fig.~\ref{fig:FriedmanLevineScaling}.  The extrapolation does not lead to a
physically meaningful limit because the boundary lengths diverge and the two
boundaries get infinitesimally close to each other, in the
$N_s\rightarrow\infty$ limit.

The extrapolation of Ref.\ \cite{friedman} arose from an incorrect adaptation
of the procedure of Ref.\ \cite{HaqueZozulyaSchoutens} which was designed for
the sphere geometry.  For the spherical case, the fixed-$l_A$ extrapolation
takes one to the well-defined limit of an infinite disk with circular $A$
region, where the boundary length indeed scales as $\sim\sqrt{l_A}$.  Also for
the disk \cite{MorrisFeder}, the fixed-$l_A$ extrapolation to
$N\rightarrow\infty$ is well-defined as the limiting situation is again an
infinite disk with constant density.  (The density is not constant in
finite-size disk simulations.)  However, for the torus with unit aspect ratio,
the fixed-$l_A$ extrapolation to $N\rightarrow\infty$ has a pathological limit
for the shape of the $A$ block, and the limiting entanglement entropy has no
reason to scale as $\sim\sqrt{l_A}$.  A plot of the extrapolated $S_A$ versus
$\sqrt{l_A}$ thus has no obvious connection to the entropy scaling as a
function of the boundary length, or to the definition of $\gamma$ as
formulated in Refs.\ \cite{KitaevPreskill_PRL06,LevinWen_PRL06}.

The idea of obtaining entanglement entropy scaling through varying discrete torus
circumferences has been employed in Ref.\ \cite{FurukawaMisguich_PRB07} for
the dimer model on the triangular lattice.  The details are quite different
from the FQH case.
It is possible that some of the ideas developed here for the FQH context might
be transferred fruitfully to numerical work on dimer models or other lattice
models.  Ref.\ \cite{integer} has studied entanglement of \emph{integer}
quantum Halls states on a torus, between true spatial partitions rather than
orbital partitions.  The orbital partitioning entanglement is zero for integer
quantum Hall states because they are product states in the orbital basis.

Since Ref.\ \cite{HaqueZozulyaSchoutens, HaqueZozulyaSchoutens2} reported
entanglements of the same Laughlin states on a different geometry, it is
interesting to compare the magnitudes of the entanglement entropy.  The data
tabulated in Ref.\ \cite{HaqueZozulyaSchoutens2} for the $\nu=1/3$ state
yields a value of $\alpha\sim0.15$, which is close to that obtained from the
torus entanglement data reported in this work.  While this is not unexpected,
it has several instructive implications.  First, it can be regarded as
additional evidence that orbital partitioning entanglement is a good
approximation to spatial partitioning entanglement.  Second, it shows that the
entanglement entropy contributions from the two edges simply add for a block
with two edges compared to one edge.

In addition, the fact that the sphere and the torus have the same
``entanglement entropy density'' per unit boundary length, allows us to
compare the difficulty of DMRG simulations on spherical and toroidal
geometries based on considerations of linear sizes alone.  Conventional
wisdom might be that the torus is significantly more difficult to treat using
DMRG, because it is a system with periodic boundary conditions while the sphere
is more analogous to a lattice system with open boundary conditions.  However,
the torus has the same block boundary length everywhere ($2L_1 =
2\sqrt{2{\pi}N_s}$ for the hardest case of unit aspect ratio), while the block boundary on a
sphere varies.  For DMRG on a spherical geometry, the dominant reduced entropy
contributions comes from the equator region, where the block boundary is
$L=\pi \sqrt{2 N_s}$.  The torus boundary is only a factor $2/\sqrt{\pi}\approx1.13$
larger than the sphere case, rather than being twice as large. Moreover, from the 
two edges one benefits twice the (negative) contribution from the topological part 
of the entanglement entropy (in contrast to a single contribution on the sphere). 
We therefore infer that DMRG on the torus geometry is not as drastically more difficult
compared to the sphere case, as would be suggested by the argument of two
block boundaries versus one.

The entanglement entropy $S$ in a bipartition of a quantum state puts 
a lower bound on the number of states $m$ to be kept in an accurate DMRG simulation
through the relation $m \sim \exp(S)$. The computational complexity of DMRG being 
polynomial in $m$, the scaling of $S$ with system geometry is thus of primary importance.
In the simplest cases, {\it e.g.}, one dimensional gapped quantum systems with local interactions,
the entropy $S$ does not depend on the block length, enabling DMRG simulations for 
basically infinite systems at constant $m$. Torus FQH simulations at {\em constant} $L_1$~\cite{bk03} also
belong to this tractable class. If one is however interested in describing true  bulk FQH
systems at fixed aspect ratio ($L_1/L_2=\mathrm{const}$), then entropy $S$ will scale linearly 
with $L_1 \propto \sqrt{N_s}$. This translates into $m \propto \exp[ \mathrm{const} \times \sqrt{N_s}]$,
i.e., accurate DMRG simulations for bulk FQH states scale exponentially in the physical 
width, similar to 2D lattice models~\cite{dmrgrev}.

The present work opens up a number of directions deserving exploration.  Our
analysis provides a way to decide on whether or not available wavefunction
sizes provide access to the entanglement scaling regime.  Thus, as bigger
wavefunctions become available, our analysis can be applied directly to obtain
better calculations of the topological entanglement entropy $\gamma$.
Eventually, this type of calculation could become a standard tool for
diagnosing unknown or poorly-understood FQH states.  Another obvious direction
is the study of entanglement scaling through geometric deformations at other
fractions and more complicated states.  It may also be interesting to try to
devise continuous geometric tuning parameters for other geometries,
\emph{e.g.}, one may consider ellipsoidal geometries as deformations of the
sphere, although setting up the Landau level problem in such geometries is not
straightforward or convenient.  It is possible that a combination of various
deformation considerations might lead to further refined procedures for
estimating entanglement scaling and $\gamma$.

\section{Acknowledgments}

We thank Kareljan Schoutens for suggesting entanglement calculations on the torus 
with varying geometry, 
and Juha Suorsa for collaboration on related topics.
MH thanks Nicolas Regnault for related discussions. 
We acknowledge MPG RZ Garching and ZIH TU Dresden for allocation of computing time.

\end{document}